\documentclass[12pt,]{article}
\usepackage{lmodern}
\usepackage{amssymb,amsmath}
\usepackage{ifxetex,ifluatex}
\usepackage{fixltx2e} 
\ifnum 0\ifxetex 1\fi\ifluatex 1\fi=0 
  \usepackage[T1]{fontenc}
  \usepackage[utf8]{inputenc}
\else 
  \ifxetex
    \usepackage{mathspec}
  \else
    \usepackage{fontspec}
  \fi
  \defaultfontfeatures{Ligatures=TeX,Scale=MatchLowercase}
\fi
\IfFileExists{upquote.sty}{\usepackage{upquote}}{}
\IfFileExists{microtype.sty}{%
\usepackage{microtype}
\UseMicrotypeSet[protrusion]{basicmath} 
}{}
\usepackage[margin=1in]{geometry}
\usepackage{hyperref}
\hypersetup{unicode=true,
            pdftitle={The Propensity to Cycle Tool: An open source online system for sustainable transport planning},
            pdfauthor={Robin Lovelace (University of Leeds); Anna Goodman (London School of Hygiene and Tropical Medicine); Rachel Aldred (University of Westminster); Nikolai Berkoff (independent web developer); Ali Abbas (University of Cambridge); James Woodcock (University of Cambridge)},
            pdfborder={0 0 0},
            breaklinks=true}
\usepackage{graphicx,grffile}
\makeatletter
\def\maxwidth{\ifdim\Gin@nat@width>\linewidth\linewidth\else\Gin@nat@width\fi}
\def\maxheight{\ifdim\Gin@nat@height>\textheight\textheight\else\Gin@nat@height\fi}
\makeatother
\setkeys{Gin}{width=\maxwidth,height=\maxheight,keepaspectratio}
\IfFileExists{parskip.sty}{%
\usepackage{parskip}
}{
\setlength{\parindent}{0pt}
\setlength{\parskip}{6pt plus 2pt minus 1pt}
}
\providecommand{\tightlist}{%
  \setlength{\itemsep}{0pt}\setlength{\parskip}{0pt}}
\ifx\paragraph\undefined\else
\let\oldparagraph\paragraph
\renewcommand{\paragraph}[1]{\oldparagraph{#1}\mbox{}}
\fi
\ifx\subparagraph\undefined\else
\let\oldsubparagraph\subparagraph
\renewcommand{\subparagraph}[1]{\oldsubparagraph{#1}\mbox{}}
\fi

\let\rmarkdownfootnote\footnote%
\def\footnote{\protect\rmarkdownfootnote}

\usepackage{titling}

  \title{The Propensity to Cycle Tool: An open source online system for
sustainable transport planning}
  \pretitle{\vspace{\droptitle}\centering\huge}
  \posttitle{\par}
  \author{Robin Lovelace (University of Leeds) \\ Anna Goodman (London School of Hygiene and Tropical Medicine) \\ Rachel Aldred (University of Westminster) \\ Nikolai Berkoff (independent web developer) \\ Ali Abbas (University of Cambridge) \\ James Woodcock (University of Cambridge)}
  \preauthor{\centering\large\emph}
  \postauthor{\par}
  \date{}
  \predate{}\postdate{}

\usepackage{amsmath} \usepackage{longtable} \usepackage{booktabs}
\usepackage{setspace}  \usepackage{lscape}
\usepackage{rotating} \usepackage{tabularx} \usepackage{tabu}
\newcommand{\blandscape}{\begin{landscape}}
\newcommand{\elandscape}{\end{landscape}} \usepackage{caption}
\begin{document}
\maketitle

\section*{Abstract}\label{abstract}
\addcontentsline{toc}{section}{Abstract}

Getting people cycling is an increasingly common objective in transport
planning institutions worldwide. A growing evidence base indicates that
high quality infrastructure can boost local cycling rates. Yet for
infrastructure and other cycling measures to be effective, it is
important to intervene in the right places, such as along `desire lines'
of high latent demand. This creates the need for tools and methods to
help answer the question `where to build?'. Following a brief review of
the policy and research context related to this question, this paper
describes the design, features and potential applications of such a
tool. The Propensity to Cycle Tool (PCT) is an online, interactive
planning support system which was initially developed to explore and map
cycling potential across England (see
\href{http://pct.bike/}{www.pct.bike}). Based on origin-destination
data, it models and visualises cycling levels at area, desire line,
route and route network levels, for current levels of cycling, and for
scenario-based `cycling futures'. Four scenarios are presented,
including `Go Dutch' and `Ebikes', which explore what would happen if
English people cycled as much as Dutch people and the potential impact
of electric cycles on cycling uptake. The cost effectiveness of
investment depends not only on the number of additional trips cycled,
but on wider impacts such as health and carbon benefits. The PCT reports
these at area, desire line, and route level for each scenario. The PCT
is open source, facilitating the creation of additional scenarios and
its deployment in new contexts. We conclude that the PCT illustrates the
potential of online tools to inform transport decisions and raises the
wider issue of how models should be used in transport planning.

\section{Introduction}\label{introduction}

Cycling can play an important role in creating sustainable and equitable
transport systems. Cycling already provides reliable, healthy,
affordable, and convenient mobility to millions of people each day
(Komanoff, 2004) and is one of the fastest growing modes of transport in
cities such as London, New York and Barcelona (Fishman, 2016). There is
mounting evidence about the external costs of car-dominated transport
systems (e.g. Han and Hayashi, 2008; Shergold et al., 2012), and the
benefits of cycling (De Nazelle et al., 2011; Oja et al., 2011; Tainio
et al., 2016). In this context there is growing interest, and in some
cases substantial investment, in cycling infrastructure, including in
countries with historically low rates of cycling.

Providing high-quality infrastructure can play a key role in promoting
cycling uptake (Parkin, 2012). Off-road cycle paths, for example, have
been found to be associated with an uptake of cycling for commuting
(Heinen et al., 2015). Overall there is growing evidence linking cycling
infrastructure to higher rates of cycling (Buehler and Dill, 2016). But
where should this infrastructure be built? The central purpose of this
paper is highlight the potential of online, interactive and publicly
accessible tools to improve the provision of locally specific evidence
about where cycling has the greatest potential to grow for transport
planning. It does so with reference to the Propensity to Cycle Tool
(PCT), an online planning support system funded by the UK's Department
for Transport to map cycling potential (Department for Transport, 2015).
The PCT is open source. This differs from other tools for assessing
cycling potential (see the next section) and has wider implications for
how models can and should be used in the transport decision making
process (see the Discussion).

\section{The Propensity to Cycle Tool in
context}\label{the-propensity-to-cycle-tool-in-context}

The PCT was developed alongside two branches of academic research: a)
methodological developments for estimating cycling potential and b)
Planning Support Systems (PSS). The subsequent overview of this policy
and academic landscape places the PCT in its wider context and explains
its key features.

\subsection{The policy context}\label{the-policy-context}

A number of factors influence the attractiveness of cycling for everyday
trips (Parkin, 2015; Pucher et al., 2010). However, the intervention
that has received the most attention has been the construction of new
cycle paths. In the UK context, devolved transport budgets mean that
local authorities have some control over the design and implementation
of cycling networks. While much of the road network budget allocated
centrally, through Highways England, cycle paths tend to be commissioned
and designed locally due to funding streams such as the Local
Sustainable Transport Fund, described in the context of modal shift to
cycling by Chatterjee et al. (2013). This, and the shorter distances and
more flexible routes (e.g.~through parks and along disused railways)
that cycle paths can take increases the importance of high resolution
map-based tools for cycling compared with motorised modes. Local
transport budgets also increase demand for local cycling targets that
open source tools such as the PCT could enable, as described in the
Discussion.

Planning new cycle paths requires many decisions to be made, including
in relation to the width (Pikora et al., 2002; Wegman, 1979), quality
(Heath et al., 2006), directness (CROW, 2007) and geographic location of
the paths. Yet while much guidance has been produced regarding physical
design (e.g. Transport for London, 2015; Welsh Government, 2014), little
work has explicitly tackled the question of where high quality
infrastructure should be built (Aultman-Hall et al., 1997; Larsen et
al., 2013; Minikel, 2012). Within this policy context, the PCT focuses
explicitly on the question of \emph{where} to build rather than
\emph{what} to build, although it does provide evidence on potential
capacity requirements across the route network.

\subsection{Research into cycling
potential}\label{research-into-cycling-potential}

There is an emerging literature exploring cycling rates and potential
rates of growth. Much of this work is practitioner-led, as detailed in
an excellent overview of developments in the USA (Kuzmyak et al., 2014).
With the notable exceptions of Larsen et al. (2013) and Zhang et al.
(2014), the methods reviewed in this section are not simultaneously
\emph{route-based}, \emph{systematic}, \emph{quantitative} and
\emph{scalable}, some of the key features of the PCT. The work can be
classified in a number of ways, including by the main mechanism of the
tool (e.g.~facility demand or mode choice models), the format of its
outputs (e.g.~spreadsheet results, GIS-based map or on-line, interactive
map) or by the main level of the input data used. In these terms, the
PCT is based on a mode choice model (representing the shift to cycling
from other modes) operating primarily at the origin-destination level,
which provides outputs in an online, interactive map. For research into
cycling potential (and the resulting models) to be useful for local
transport planners, their spatial scale must coincide with those over
which the planning process has some control. For this reason we focus on
scale as the primary way of categorising research into cycling
potential, highlighting the following prominent levels (the PCT best
fits into the third).

\begin{itemize}
\item
  Area-based measures are based primarily on data at the level of
  administrative zones, which will vary depending on context. Outputs
  from these measures can assist with the location of site-specific
  transport infrastructure such as cycle parking.
\item
  Individual-based measures are based on survey data, typically a
  household travel survey. These are not always geographically specific
  and tend to be used to identify and categorise demographic groups in
  relation to cycling, such as near-market or as warranting tailored
  interventions, such as targeted cycle training schemes.
\item
  Route-based measures use origin-destination data which can be used to
  create `desire lines' and (using route allocation) estimates of
  existing and potential demand at each point on the road network.
\end{itemize}

This work is reviewed in relation to the PCT below and summarised in
Table 1.

Parkin et al. (2008) presented an area-based measure of cycling
potential using regression model to estimate the proportion of commuter
trips cycled across wards in England and Wales. Factors associated with
lower levels of cycling included road defects, high rainfall, hills and
a higher proportion of ethnic minority and low-income inhabitants.
Parkin et al. concluded that policy makers must engage with a mixture of
physical and social barriers to promote cycling effectively, with the
implication that some areas have lower barriers to cycling --- and hence
higher propensity to cycle --- than others.

Zhang et al. (2014) created an individual-based model of cycling
potential to prioritise where to build cycle paths to ``achieve maximum
impacts early on''. The outputs of this model were aggregated to the
level of 67 statistical zones in the study area of Belo Horizonte,
Brazil, and used to generate a `usage intensity index' for potential
cycle paths. This, combined with survey data on cyclists' stated
preferences on whether people would cycle were infrastructure provided
along particular routes and origin-destination data on travel to work,
was used to rank key routes in the city in terms of their cycling
potential.

Larsen et al. (2013) created an area-based `prioritization index', for
Montreal, Canada. This was based on four variables: the area's current
level of cycling, its cycling potential (estimated based on the shortest
path between the origin and destination of short car trips from a travel
survey), the number of injuries to cyclists, and locations prioritised
by current cyclists for improvement (Larsen et al., 2013). These four
were aggregated to the level of evenly spread cells covering the study
area. The resulting heat map was used to recommend the construction or
upgrade of cycle paths on specific roads.

Although the methods presented in preceding three paragraphs were
developed in an academic context, most of the usable \emph{tools} for
estimating cycling potential are practitioner-led. Many of these may
never have been documented in the academic literature, so the subsequent
overview should not be regarded as comprehensive (see Kuzmyak et al.,
2014 and subsequent documents that cite this report).

The Analysis of Cycling Potential (ACP) tool was developed by Transport
for London (2010) to produce heat maps of cycling potential across
London, for all trip purposes. It is based a mode choice model of
likelihood of trips being cycled and uses the characteristics of
observed trips (e.g.~time of day, characteristics of the traveller,
distance) as the main input dataset. The results of the ACP have
informed local cycling schemes, such as where to build new cycle hire
stations but does not use origin-destination data so is less relevant
for route-based interventions.

A more localised approach is the Permeability Assessment Tool (PAT),
which was developed by a transport consultancy Payne (2014). The PAT is
based on the concept of `filtered permeability', which means providing a
more direct route to people cycling than driving (Melia, 2015). The PAT
works by combining geographical data, including the location of popular
destinations and existing transport infrastructure, with on-site audit
data of areas that have been short-listed. Unlike the prioritisation
index of Larsen et al. (2013), which is primarily aimed at informing a
city-wide strategic cycling network, the results of the PAT are designed
to guide smaller, site specific interventions such as `contraflow' paths
and cyclist priority traffic signals.

The Santa Monica model is an example of a direct \emph{facility demand}
model, which are based on ``observed counts and context-driven
regression models'' (Kuzmyak et al., 2014). The model uses environmental
explanatory variables such as employment density and speed limits of
surrounding roads to estimate the dependent variable: walking and
cycling activity (e.g.~as recorded by screenline count devices). By
adjusting regression parameters (e.g.~based on areas that have
experienced growth in cycling), such models can be used to ``forecast
demand levels for walk or bike at a point or intersection level'' and
``evaluate and prioritize projects'' (ibid., p.~75).

\begin{table}[ht]
\centering
\caption{Summary of tools and methods to explore the geographical distribution of cycling potential. Levels of analysis refer to whether data is analysed at point (P), area (A), origin-destination (OD), route (R), route network (RN) or individual (I) levels.} 
\begin{tabularx}{\textwidth}{p{2.5cm}|p{2cm}p{2cm}XXXp{1.8cm}}
 Tool & Scale & Coverage & Public access & Format of output & Levels of analysis & Software licence \\ 
  \midrule
Propensity to Cycle Tool & National & England & Yes & Online map & A, OD, R, RN & Open source \\ 
   \midrule
Prioritization Index & City & Montreal & No & GIS-based & P, A, R & Proprietary \\ 
   \midrule
Permeability Assessment Tool & Local & Parts of Dublin & No & GIS-based & A, OD, R & Proprietary \\ 
   \midrule
Usage intensity index & City & Belo Horizonte & No & GIS-based & A, OD, R, I & Proprietary \\ 
   \midrule
Bicycle share model & National & England, Wales & No & Static & A, R & Unknown \\ 
   \midrule
Cycling Potential Tool & City & London & No & Static & A, I & Unknown \\ 
   \midrule
Santa Monica model & City & Santa Monica & No & Static & P, OD, A & Unknown \\ 
   \bottomrule
\end{tabularx}
\end{table}

\subsection{Planning support systems}\label{planning-support-systems}

The methods and tools for estimating cycling potential outlined in Table
1 were generally created with only a single study region in mind. The
benefit of this is that they can respond context-specific to
practitioner and policy needs. However, the PCT aims to provide a
\emph{generalisable} and \emph{scalable} tool, in the tradition of
Planning Support Systems (PSS).

PSS were developed to encourage evidence-based policy in land-use
planning (Klosterman, 1999). The application of PSS to transport
planning has been more recent, with a goal of ``systematically
{[}introducing{]} relevant (spatial) information to a specific process
of related planning actions'' (Brömmelstroet and Bertolini, 2008). The
PCT is systematic in its use of national data for all parts of the study
region (in this case England) and relates to a specific planning process
--- the creation of new and enhancement of existing cycle
infrastructure.

PSS typically work by presenting evidence about the characteristics and
needs of the study region in an interactive map. A central objective is
to visualise alternative scenarios of change, and explore their
potential impacts. The results of traditional scenario-based models are
usually not locally specific (Lovelace et al., 2011; McCollum and Yang,
2009; Woodcock et al., 2009). Online PSS can overcome this issue by
using interactive maps (Pettit et al., 2013). The emergence of libraries
for web mapping (Haklay et al., 2008) has facilitated online PSS,
offering the potential for public access to the planning process.
Transparency is further enhanced by making PSS open source, in-line with
a growing trend in transport modelling (Borning et al., 2008; Novosel et
al., 2015; Tamminga et al., 2012). In these ways, PSS can make evidence
for transport planning more widely available, and tackle the issue that
transport models are often seen as `black boxes', closed to public
scrutiny (Golub et al., 2013).

With reference to a publicly accessible online map-based PSS for
strategic land use (and to a lesser extent transport) planning,
Kahila-Tani et al. (2016) provide guidance on the wider public
participation process: it should be \emph{transparent}, \emph{influence}
real world decisions, be \emph{representative} of citizens and allow
\emph{independent participation}. Although these criteria depend
primarily on \emph{how} the tool is deployed in practice (for example
with regards to training resources and workshops facilitated by planning
authorities to ensure that the tool the tool is used effectively), which
are outside the scope of the current paper, the PCT certainly encourages
transparency through its use of open data and open source software and
allows independent participation through its publicly accessible online
interface.

\subsection{National context and features of the Propensity to Cycle
Tool}\label{national-context-and-features-of-the-propensity-to-cycle-tool}

In addition to the international policy and academic context, the PCT
was influenced by the national context. It was commissioned by the UK's
Department for Transport to identify ``parts of {[}England{]} with the
greatest propensity to cycle'' (Department for Transport, 2015). Thus
the aim was not to produce a full transport demand or land use model,
but to provide an evidence base to prioritise where to create and
improve cycling infrastructure based on scenarios of change.

Local and national cycling targets are often based on a target mode
share by a given date.\footnote{The local target in Bristol, for
  example, is for
  \href{http://www.bristolpost.co.uk/Bristol-City-Council-s-pound-35million-plan/story-21341172-detail/story.html}{20\%
  of commuter trips to be cycled by 2020}.
  \href{http://cycling.tfgm.com/Documents/CCAG2-Executive-Summary.pdf}{Manchester}
  (10\% by 2025),
  \href{https://www.derbyshire.gov.uk/images/2015-07-07\%20Item\%207l\%20Cycle\%20Plan_tcm44-267216.pdf}{Derbyshire}
  (to double the number of people cycling by 2025) and
  \href{http://content.tfl.gov.uk/gla-mayors-cycle-vision-2013.pdf}{London}
  (to `double cycling' by 2025) provide further examples of local
  ambitious time-bound cycling targets.} However, there is little
evidence about what this might mean for cycling volumes along specific
routes. The PCT tackles this issue by estimating rate of cycling locally
under different scenarios and presenting the results on an interactive
map. Its key features include:

\begin{itemize}
\item
  Estimation of cycling potential at area, `desire line' and route
  network levels.
\item
  Route-allocation of OD (origin-destination) pairs by a routing
  algorithm specifically developed for cycling. This was done by
  CycleStreets.net, a routing service developed by cyclists, for
  cyclists.
\item
  Visualisation of outputs at multiple geographic levels. The
  interactive map enables users to examine cycling potential at a very
  local level (e.g.~just a few streets) or at a more regional level
  (e.g.~across a large metropolitan area).
\item
  Public accessibility of results and code. The tool is freely available
  online and developers are encouraged to modify the PCT (e.g.~to create
  alternative scenarios) by provision of the source code underlying the
  PCT under the open source AGP License.
\item
  The presentation of estimated health economic and carbon impacts under
  future scenarios, providing an evidence base that could be used in
  business cases for investment.
\end{itemize}

As with any tool, the PCT's utility depends on people knowing how to use
it. For that reason training materials and a user manual are being
developed to show how the tool can be used (see the `Manual' tab in
Figure 3 and \href{http://pct.bike/manual.html}{pct.bike/manual.html}).

\section{Data and methods}\label{data-and-methods}

This section describes the data and methods that generate the scenario
data for the PCT, as summarised in Figure 1. The details of the model,
which models the proportion of trips made by cycling per OD pair as a
function of hilliness and distance, are described the Appendix. The
scenario-generation method is not included in the main text of the paper
because it is the most context-specific aspect of the PCT: as outlined
in the Discussion, we envision future uses of the PCT using different
model parameters and even different functional forms relating distance,
hilliness and other variables to the level of cycling, to generate
scenarios of use for new contexts beyond the English case study
described here. Central to the PCT approach is origin-destination (OD)
data recording the travel flow between administrative zones.\footnote{The
  size and uniformity of these depend on the country in question. In the
  UK the primary areal units for statistical data are output areas (OA),
  lower layer super output areas (LSOA) and middle layer super output
  areas (MSOA) (see
  \href{http://webarchive.nationalarchives.gov.uk/20160105160709/http://www.ons.gov.uk/ons/guide-method/geography/beginner-s-guide/census/super-output-areas--soas-/index.html}{webarchive.nationalarchives.gov.uk}).
  The version of the PCT presented in this paper operates at the MSOA
  level.} Combined with geographical data on the coordinates of the
population-weighted centroid of each zones, these can be represented as
straight `desire lines' or as routes allocated to the transport network.

\begin{figure}
\includegraphics[width=1\linewidth]{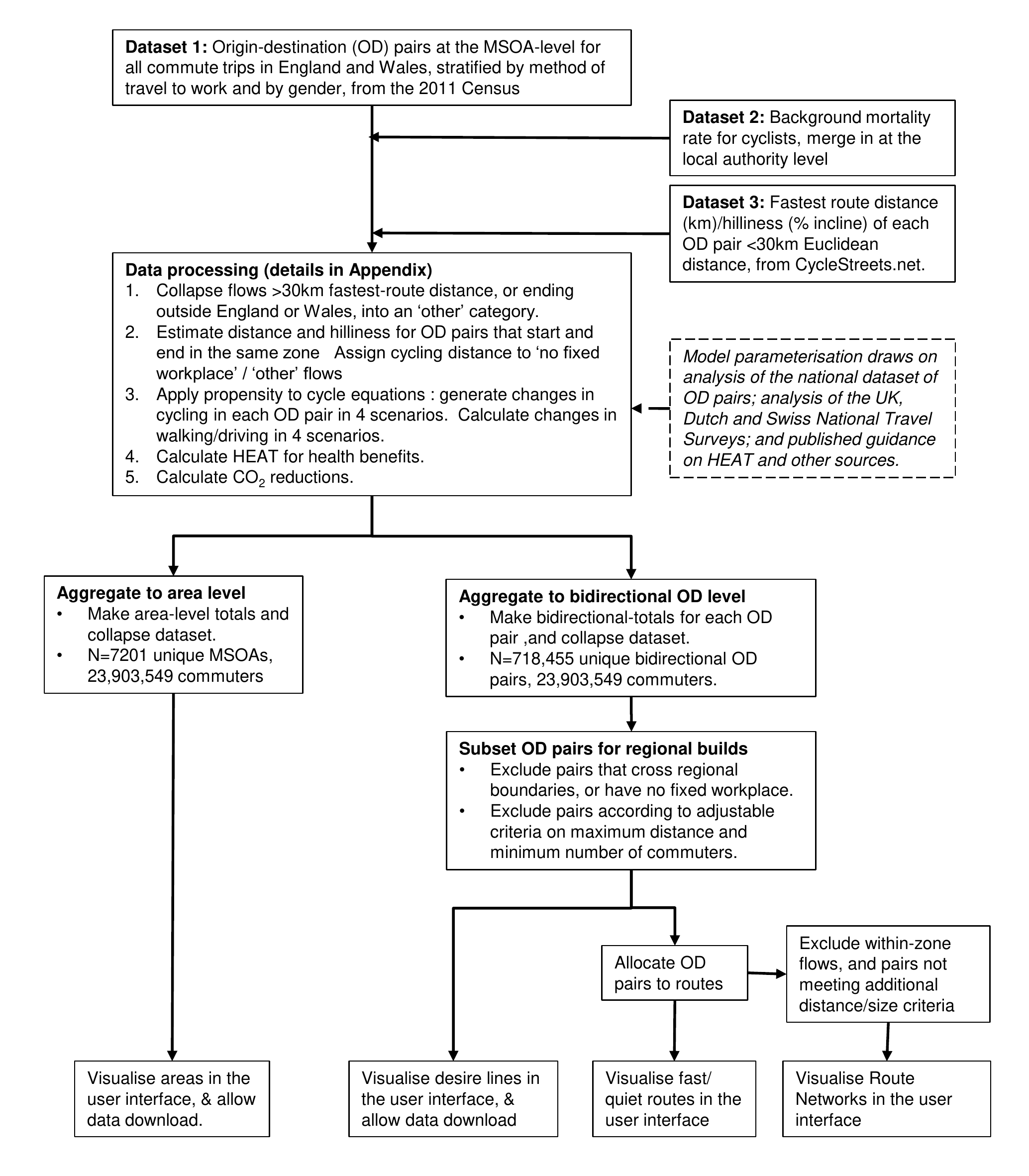} \caption{Flow diagram illustrating the input data and processing steps used to create the input data used by the PCT. The abbreviations are as follows: HEAT = Health Economic Assessment Tool, OD pair = origin-destination pair, MSOA = Middle-Layer Super Output Area}\label{fig:unnamed-chunk-8}
\end{figure}

\begin{verbatim}
## [1] FALSE
\end{verbatim}

\subsection{Processing OD data}\label{processing-od-data}

The central input dataset was a table of origin-destination (OD) pairs
from the 2011 Census. This was loaded from open access file
\texttt{wu03ew\_v2.csv}, provided by the
\href{http://wicid.ukdataservice.ac.uk/}{UK Data Service}. This captures
the number of commuters travelling between Middle Super Output Area
zones (MSOAs, average commuter population: 3300), by mode of travel (see
Table 2). This dataset was derived from responses to the following
questions in the \href{http://www.ons.gov.uk/census/}{2011 Census for
England and Wales}: ``In your main job, what is the address of your
workplace?'' (question 40) and ``How do you usually travel to work?
(Tick one box only, for the longest part, by distance, of your usual
journey to work)'' (Question 41). This dataset was enhanced by merging
in information on the gender composition of cyclists in each OD pair
(Dataset 1 in Figure 1); data at the area level on the background
mortality rate (Dataset 2); and data at the OD pair-level on route
distance (km) and hilliness (average gradient, as a percentage) (Dataset
3). OD data was assigned to the transport network using the R package
stplanr (Lovelace et al., 2016). See the Appendix for further details.

\begin{table}[]
\centering
\caption{Sample of the OD (origin-destination) input dataset, representing the number of people who
commute from locations within and between administrative zones (MSOAs). Note `Car' refers to people who drive as their main mode of travel per OD pair, rather than people who travel to work as a passenger in a car.}
\label{my-label}
\begin{tabular}{@{}lllllll@{}}
\toprule
                                        &                                        & \multicolumn{5}{l}{Number of commuters by main mode}                                                                 \\ \midrule
\multicolumn{1}{l}{Area of residence} & \multicolumn{1}{l}{Area of workplace} & \multicolumn{1}{l}{Total} & \multicolumn{1}{l}{Cycle} & \multicolumn{1}{l}{Walk} & \multicolumn{1}{l}{Car} & \multicolumn{1}{l}{Other} \\ \midrule
\multicolumn{1}{l}{E02002361}         & \multicolumn{1}{l}{E02002361}         & \multicolumn{1}{l}{109}   & \multicolumn{1}{l}{2}     & \multicolumn{1}{l}{59}   & \multicolumn{1}{l}{39}  & \multicolumn{1}{l}{9}     \\ 
\multicolumn{1}{l}{E02002361}         & \multicolumn{1}{l}{E02002362}         & \multicolumn{1}{l}{7}     & \multicolumn{1}{l}{1}     & \multicolumn{1}{l}{0}    & \multicolumn{1}{l}{4}   & \multicolumn{1}{l}{2}     \\ 
\multicolumn{1}{l}{E02002361}         & \multicolumn{1}{l}{E02002363}         & \multicolumn{1}{l}{38}    & \multicolumn{1}{l}{0}     & \multicolumn{1}{l}{4}    & \multicolumn{1}{l}{24}  & \multicolumn{1}{l}{10}    \\ 
\multicolumn{1}{l}{E02002361}         & \multicolumn{1}{l}{E02002364}         & \multicolumn{1}{l}{15}  & \multicolumn{1}{l}{1}     & \multicolumn{1}{l}{0}    & \multicolumn{1}{l}{10}  & \multicolumn{1}{l}{4}     \\ 
\multicolumn{1}{l}{E02002361}         & \multicolumn{1}{l}{E02002366}   & 29                         & 1                          & 10                        & 11                       & 7                          \\ \bottomrule
\end{tabular}
\end{table}

\subsection{Modelling baseline propensity to
cycle}\label{modelling-baseline-propensity-to-cycle}

The starting point for generating our scenario-based `cycling futures'
was to model baseline data on cycle commuting in England and Wales. We
did this using OD data from the 2011 Census, and modelling cycling
commuting as a function of route distance and route hilliness. We did so
using logistic regression applied at the individual level, including
squared and square-root terms to capture `distance decay' --- the
non-linear impact of distance on the likelihood of cycling (Iacono et
al., 2008) --- and including terms to capture the interaction between
distance and hilliness. Model fit is illustrated in Figure 2; see the
Appendix for details and for the underlying equations. We also developed
equations to estimate commuting mode share among groups not represented
in the between-zone (`interzonal') OD data, e.g.~those commuting within
a specific MSOA (this is within-zone or `intrazonal' travel), or those
with no fixed workplace. This model of baseline propensity to cycle
formed the basis of three of the four scenarios (Government Target, Go
Dutch and Ebikes), as described in more detail in the next section.

\begin{figure}
\includegraphics[width=1\linewidth]{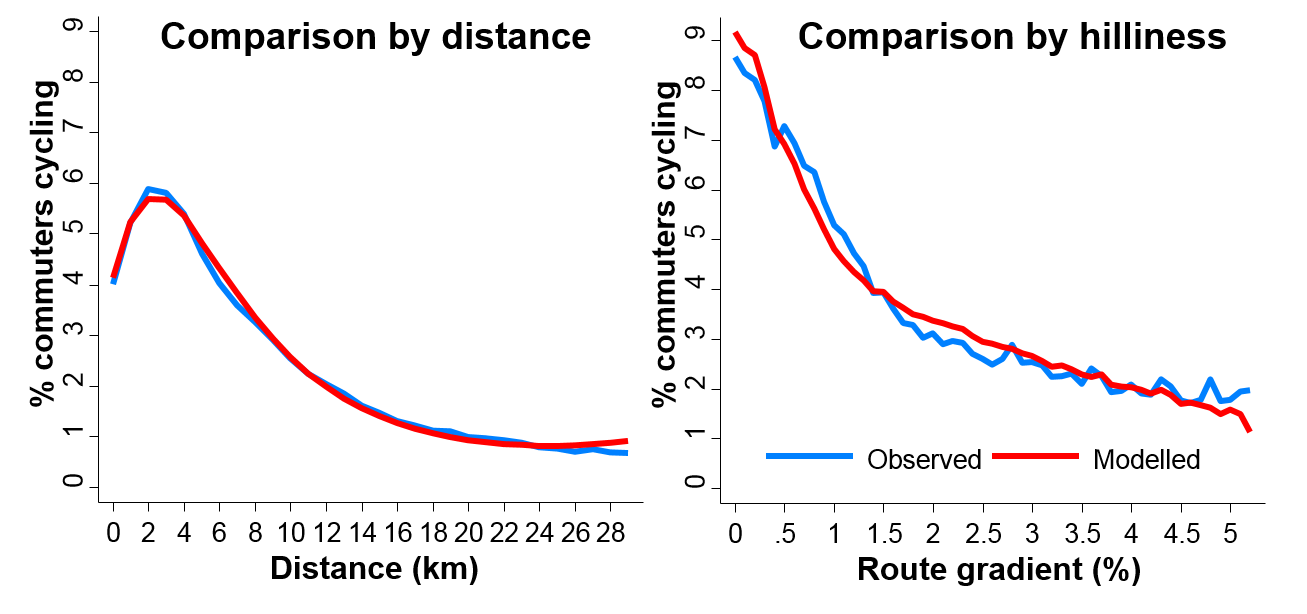} \caption{The relationship between distance (left) and hilliness (right) and cycling mode share based on data from the 2011 Census for England and Wales. The plots show actual (blue) vs predicted (red) prevalence of cycling to work among commuters travelling <30km to work.}\label{fig:unnamed-chunk-9}
\end{figure}

\begin{verbatim}
## [1] FALSE
\end{verbatim}

\subsection{Scenarios of cycling
uptake}\label{scenarios-of-cycling-uptake}

Four scenarios were developed to a range of explore cycling futures.
These can be framed in terms of the removal of different
infrastructural, cultural and technological barriers that currently
prevent cycling being the natural mode of choice for trips of short to
medium distances. They are not predictions of the future. They are
snapshots indicating how the spatial distribution of cycling may shift
as cycling grows based on current travel patterns. At a national level,
the first two could be seen as shorter-term and the second two more
ambitious. The choice of scenarios was informed by a government target
to double the number of cycle trips and evidence from overseas about
which trips \emph{could} be made by cycling. Summaries of the four
scenarios are as follows (see the Appendix for full details):

\begin{itemize}
\item
  Government Target. This scenario represents a doubling of the level of
  cycling, in line with the government's target to double the number of
  `stages' (trips of a single mode) cycled by 2025 (Department for
  Transport, 2014). Although substantial in relative terms, the rate of
  cycling under this scenario (rising from 3\% to 6\% of English
  commuters) remains low compared with countries such as the Netherlands
  and Denmark. This scenario was generated by adding together a) the
  observed number of cyclists in each OD pair in the 2011 Census, and b)
  the modelled number of cyclists, as estimated using the baseline
  propensity to cycle equations described in the previous section. The
  result is that cycling overall doubles at the national level, but at
  the local level this growth is not uniform, in absolute or relative
  terms. Areas with many short, flat trips and a below-average current
  rate of cycling are projected to more than double. Conversely, areas
  with above-average levels of cycling and many long-distance hilly
  commuter routes will experience less than a doubling.
\item
  Gender Equality. This scenario illustrates the increase in cycling
  that would result if women were as likely as men to cycle a given
  trip. Specifically, the scenario sets the proportion of female cycle
  commuters to be equal to the current proportion of males in each OD
  pair. The scenario is based on the observation that in places where
  cycling accounts for a high proportion of personal travel, women cycle
  at least as much as men (Aldred et al., 2016; Pucher et al., 2010).
  This scenario has the greatest relative impact in areas where the rate
  of cycling is highly gender-unequal. 
\item
  Go Dutch. While the Government Target and Gender equality scenarios
  model relatively modest increases in cycle commuting, Go Dutch
  represents what would happen if English people were as likely as Dutch
  people to cycle a trip of a given distance and level of hilliness.
  This scenario thereby captures the proportion of commuters that would
  be expected to cycle if all areas of England had the same
  infrastructure and cycling culture as the Netherlands (but retained
  their hilliness and commute distance patterns). This scenario was
  generated by taking baseline propensity to cycle (see previous
  section), and applying Dutch scaling factors --- parameters which
  increase the proportion of trips cycled above the baseline level of
  cycling per OD pair. These scaling factors were calculated through
  analysis of the British and Dutch National Travel Surveys. We
  parameterised the scaling factors as one fixed parameter (the main
  effect) plus one distance-dependent parameter (parametrised as an
  interaction between Dutch status and trip distance), to take into
  account the fact that the ``Dutch multiplier'' is greater for shorter
  trips compared to longer trips. Note that the scenario level of
  cycling under Go Dutch is not affected by the current level of
  cycling, but is instead purely a function of the distance and
  hilliness of each OD pair --- i.e.~the two characteristics that
  determine baseline propensity to cycle.
\end{itemize}

\begin{itemize}
\tightlist
\item
  Ebikes. This scenario models the additional increase in cycling that
  would be achieved through the widespread uptake of electric cycles
  (`ebikes'). This scenario is generated by taking baseline propensity
  to cycle, applying the Dutch scaling factors described above, and then
  additionally applying Ebike scaling factors. The Ebikes scenario is
  thus currently implemented as an extension of Go Dutch but could in
  future be implemented as an extension of other scenarios. The Ebike
  scaling factors were generated through analysis of the UK, Dutch and
  Swiss National Travel Surveys, in which we estimated how much more
  likely it was that a given commute trip would be cycled by Ebike
  owners versus cyclists in general. We parameterised the scaling
  factors as varying according to trip distance and according to
  hilliness, by fitting interaction terms between these two
  characteristics and ebike ownership (Appendix Equation 1B). We did
  this to take account of the fact that electric cycles enable longer
  journeys and reduce the barrier of hills
\end{itemize}

The Ebikes scenario is thus currently implemented as an extension of Go
Dutch but could be implemented as an extension of other scenarios. The
Ebike scaling factors were generated through analysis of the English,
Dutch and Swiss National Travel Surveys, in which we estimated how much
more likely it was that a given commute trip would be cycled by Ebike
owners versus cyclists in general. We parameterised the Ebike scaling
factors as interactions with trip distance and with hilliness, to take
account of the fact that electric cycles enable longer journeys and
reduce the barrier of hills.

Additional scenarios could be developed (see Discussion). If deployed in
other settings, the PCT will likely benefit from scenarios that relate
to both the current policy context and long-term aspirations.

\subsection{Estimation of health and carbon
impacts}\label{estimation-of-health-and-carbon-impacts}

Because the cost effectiveness of cycling investments are influenced by
wider social impacts, estimated health economic and emissions impacts
are presented in the PCT.

An approach based on the World Health Organization's Health Economic
Assessment Tool (\href{http://www.heatwalkingcycling.org/}{HEAT}) was
used to estimate the number of premature deaths avoided due to increased
physical activity (Kahlmeier et al., 2014). To allow for the fact that
cycling would in some cases replace walking trips, HEAT estimates of the
increase in premature deaths due to the reduction in walking were also
included. The change in walking was estimated based on the assumption
that, within a given OD pair, all modes were equally likely to be
replaced by cycling. Thus all the non-cycling modes shown in Table 2
experienced the same relative decrease.

Trip duration was estimated as a function of the `fast' route distance
and average speed. For walking and cycling we applied the standard HEAT
approach. Ebikes are not specifically covered in HEAT Cycling but enable
faster travel and require less energy from the rider than traditional
bikes. Thus we estimated new speeds and intensity values for this mode,
giving a smaller benefit for every minute spent using Ebikes than
conventional cycles. For more details see the Appendix.

The risk of death varies by gender and increases rapidly with age. This
was accounted for using age and sex-specific mortality rates for each
local authority in England. For the baseline and Government Target
scenario the age distribution of cyclists recorded in the 2011 Census
was used. New cyclists under Go Dutch and Ebikes were assumed to have
the age-gender profile of commuter cyclists in the Netherlands. The
inclusion of age specific parameters and mode shift from walking shows
how the HEAT approach can generate nuanced health impact estimates using
publicly available data.

The net change in the number of deaths avoided for each OD pair was
estimated as the number of deaths avoided due to cycle commuting minus
the number of additional deaths due to reduced walking. Note that this
approach means that for some OD pairs where walking made up a high
proportion of trips, additional deaths were incurred. The monetary value
of the mortality impact was calculated by drawing on the standard `value
of a statistical life' used by the Department for Transport.

We also estimated the reduction in transport carbon emissions resulting
from decreased car driving in each scenario. This again relied on the
assumption that all modes were equally likely to be replaced by cycling.
The average CO\textsuperscript{2}-equivalent emission per kilometre of
car driving was taken as 0.186 kg, the 2015 value of an `average' car
(DEFRA, 2015).

\subsection{Visualisation, route allocation and network
generation}\label{visualisation-route-allocation-and-network-generation}

The data analysis and preparation stages described in the previous
sections were conducted using the national OD dataset for England as a
whole. By contrast, the stages described in this section were conducted
using a region-by-region approach. Transport decisions tend to be made
at local and regional levels (Gaffron, 2003), hence the decision to
display results on a per region basis.

Figure 3 shows the output: `desire lines' lines with attributes for each
OD pair aggregated in both directions (Chan and Suja, 2003; Tobler,
1987), and visualised as centroid to centroid `flows' (Rae, 2009; Wood
et al., 2010).

\begin{figure}
\includegraphics[width=1\linewidth]{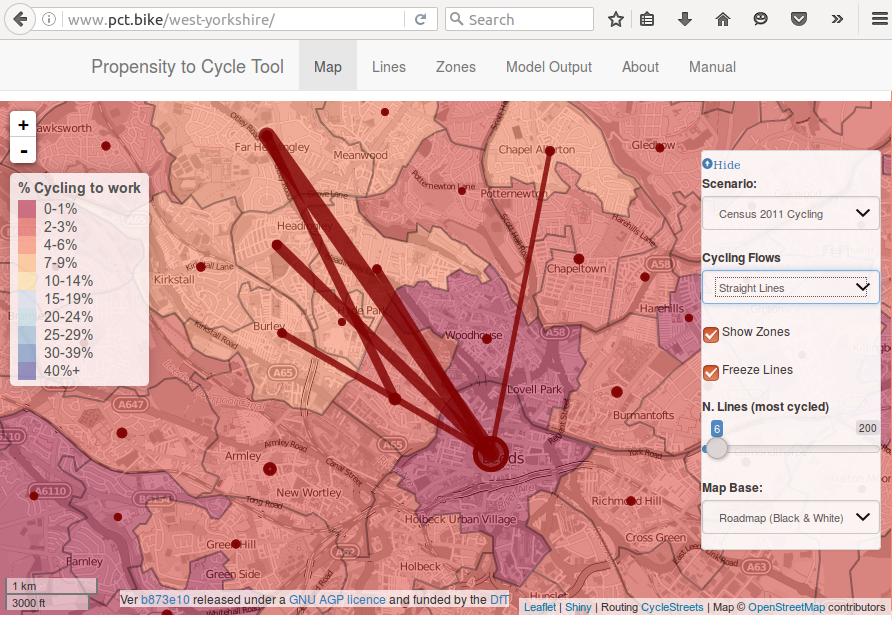} \caption{Overview of the PCT map interface, showing area and OD-level data. The zone colour represents the number of residents who cycle to work. The lines represent the top 6 most cycled commuter routes in Leeds, with width proportional to the total number of cycle trips. Population-weighted centroids are represented by circles, the diameter of which is proportional to the rate of within-zone cycling.}\label{fig:unnamed-chunk-12}
\end{figure}

\begin{verbatim}
## [1] FALSE
\end{verbatim}

Desire lines allocated to the route network are illustrated in Figure 4.
This shows two route options: the `fast' route, which represents an
estimate of the route taken by cyclists to minimise travel time and the
`quiet' route that preferentially selects smaller, quieter roads and off
road paths.

\begin{figure}
\includegraphics[width=1\linewidth]{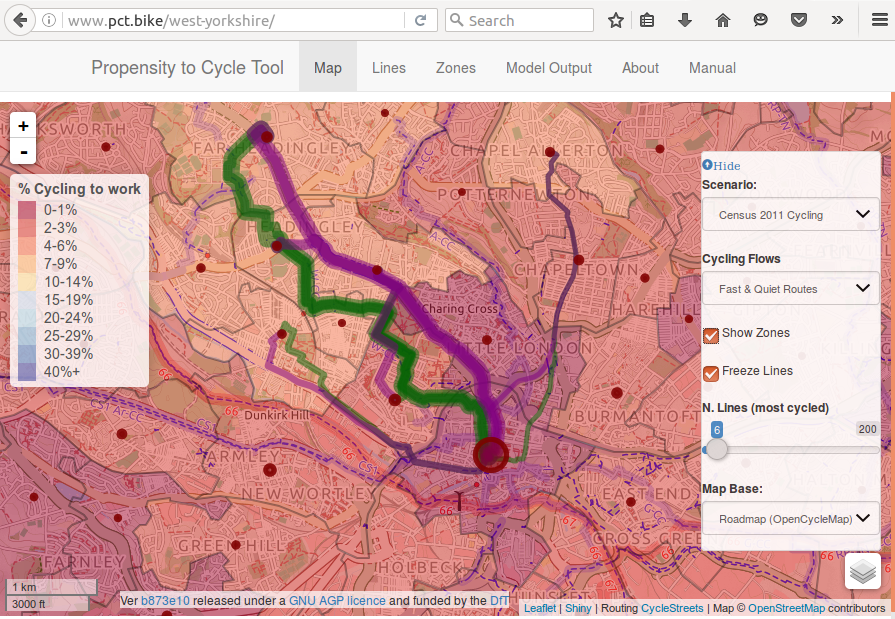} \caption{Illustration of desire lines shown in Figure 3 after they have been allocated to the road network by CycleStreets.net. Purple and green lines represent `fastest' and `quietest' routes generated by CycleStreets.net, respectively.}\label{fig:unnamed-chunk-15}
\end{figure}

\begin{verbatim}
## [1] FALSE
\end{verbatim}

Routes generated by CycleStreets.net do not necessarily represent the
paths that cyclists currently take; route choice models based on GPS
data have been developed for this purpose (Broach et al., 2012; Ehrgott
et al., 2012). Of the available routes (see
\href{https://www.cyclestreets.net/journey/help/howtouse/}{cyclestreets.net/journey/help}
for more information), the `fastest' option was used. This decision was
informed by recommendations from CROW (2007), building on evidence of
cyclists' preference direct routes.

The spatial distribution of cycling potential can be explored
interactively by selecting the `top n' routes with the highest estimated
cycling demand (see the slider entitled ``N. Lines (most cycled)'' in
Figures 3 and 4). Information about the \emph{aggregate cycling
potential} on the road network is shown in the Route Network layer.
Because the layer is the result of aggregating overlapping `fast'
routes, and summing the level of cycling for each scenario (see Figure
5), it relates to the \emph{capacity} that infrastructure may need to
handle. Cycling along Otley Road (highlighted in Figure 5), under the Go
Dutch scenario, rises from 73 to 296 commuters along a single route, but
from 546 to 1780 in the Route Network. Note that more confidence can be
placed in the relative rather than the absolute size of these numbers:
the Route Network layer excludes within-zone commuters, commuters with
no fixed workplace, and commuters working in a different region (see
Figure 1). Route Network values also omit routes due to the adjustable
selection criteria: maximum distance and minimum total numbers of
all-mode commuters per OD pair. At the time of writing these were set to
20 km Euclidean distance and 10 commuters respectively. Nationally, the
Route Network layer under these settings accounts for around two thirds
of cycle commuters.

\begin{figure}
\includegraphics[width=0.5\linewidth]{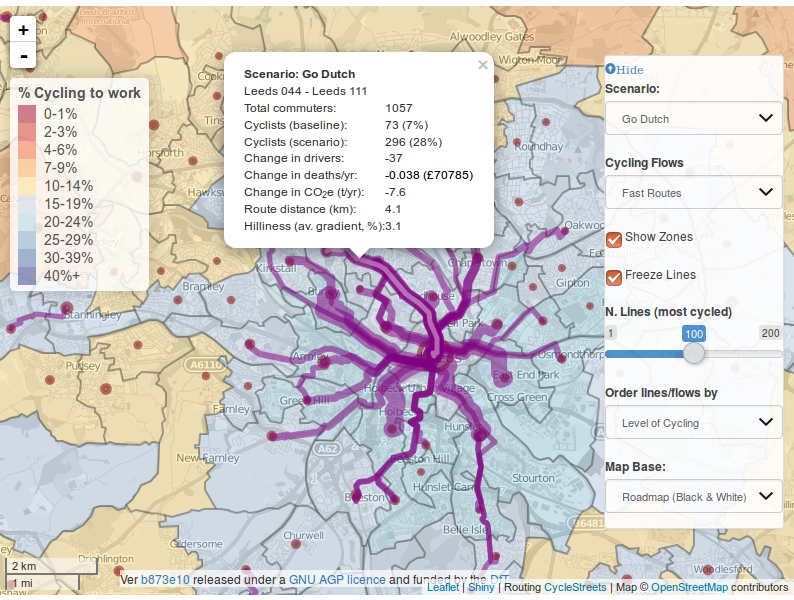} \includegraphics[width=0.5\linewidth]{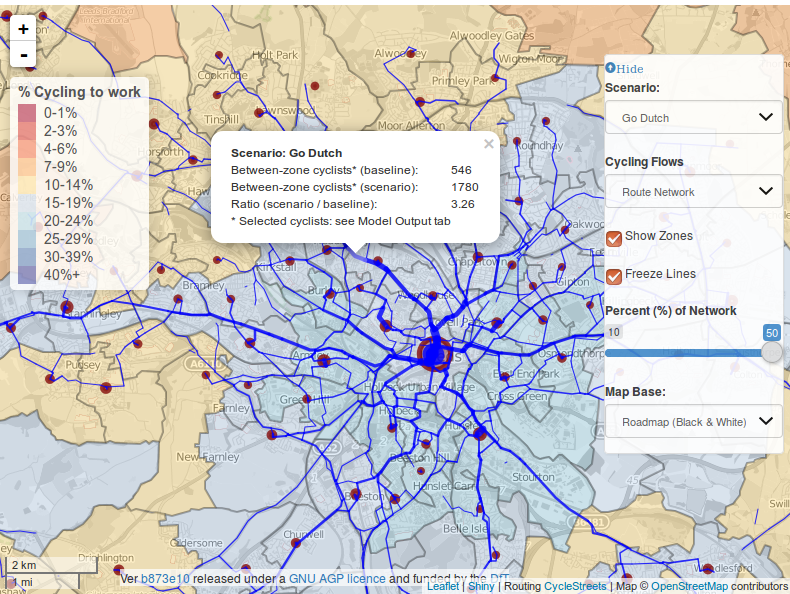} \caption{Illustration of Fast Routes map layer (left) compared with Route Network layer (right). The Route Network was was produced by aggregating all overlapping lines in the Fast Routes layer using the stplanr R package.}\label{fig:unnamed-chunk-16}
\end{figure}

\begin{verbatim}
## [1] FALSE
\end{verbatim}

\begin{verbatim}
## [1] FALSE
\end{verbatim}

\section{Outputs of the Propensity to Cycle
Tool}\label{outputs-of-the-propensity-to-cycle-tool}

This section describes and illustrates some outputs from the PCT,
alongside discussion of how these outputs could be used in transport
planning. Note that some details of the graphics in the online version
may evolve as the PCT develops.

\subsection{Model output tabs}\label{model-output-tabs}

Tabs are panels within the PCT that reveal new information when clicked
(see the top of Figure 3). Of these, the first four provide
region-specific information:

\begin{itemize}
\item
  \textbf{Map}: This interactive map is the main component of the PCT,
  and is the default tab presented to users. It shows cycling potential
  at area, desire-line, route and route network levels under different
  scenarios of change, as described throughout this paper. `Popups'
  appear when zones, desire lines or segments on the Route Network are
  clicked, presenting quantitative information about the selected
  element.
\item
  \textbf{Lines}: When lines are displayed on the interactive map, this
  tab provides raw data on a sample of the variables as a table at the
  OD pair level.
\item
  \textbf{Areas}: This tab is the equivalent of the `Lines' tab, but
  with data at the area level.
\item
  \textbf{Model output}: This tab provides key statistics, diagnostic
  plots and model results for each region. The document is produced by a
  `dynamic document' which runs embedded code for each regional dataset.
  Diagnostic plots include the distribution of cycling by trip distance
  under each scenario (see Figure 6), providing insight into local
  travel patterns and how they relate to cycling potential in the region
  overall.
\end{itemize}

\subsection{Trip distance
distributions}\label{trip-distance-distributions}

Figure 6 shows how the proportion of trips made by cycling varies as a
function of distance in two regions currently, and under the PCT's four
scenarios of change. The frequency of all mode trips by distance band
(the red lines) illustrates the two regions have different spatial
structures. Oxfordshire has a high proportion short (under 5km) trips,
helping to explain the relatively high level of cycling there. West
Yorkshire (Figure 6, left), by contrast, has a higher proportion of
longer distance commutes and a lower level of cycling than Oxford. Note
that under Go Dutch and Ebikes scenarios, regional differences in the
rate of cycling diminish, however, illustrating that these scenarios are
not influenced by the current level of cycling.

\begin{figure}
\includegraphics[width=0.5\linewidth]{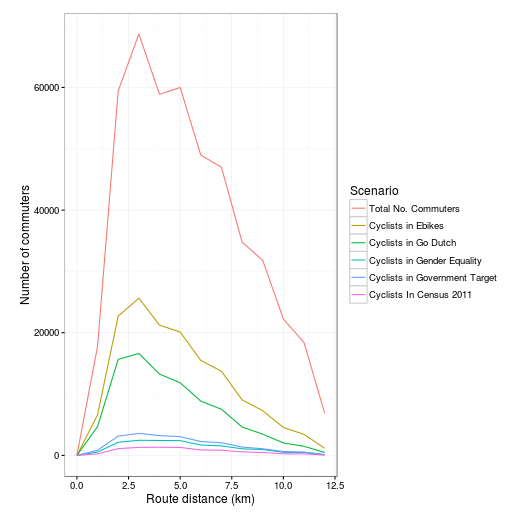} \includegraphics[width=0.5\linewidth]{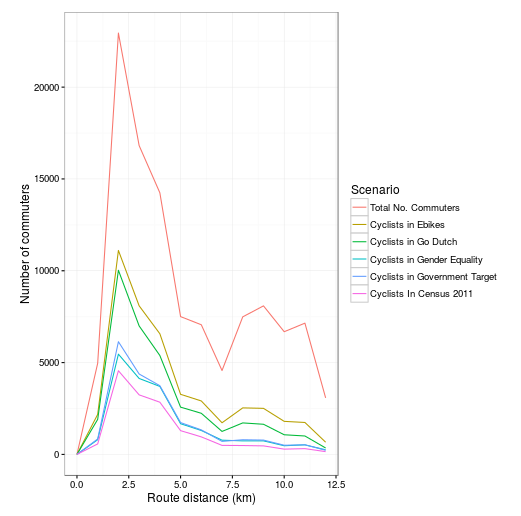} \caption{Modal share of trips made by cycling in West Yorkshire (left) and Oxfordshire (right) by distance, currently and under 4 scenarios of change.}\label{fig:unnamed-chunk-19}
\end{figure}

\begin{verbatim}
## [1] FALSE
\end{verbatim}

\begin{verbatim}
## [1] FALSE
\end{verbatim}

\subsection{The shifting spatial distribution of cycling and associated
impacts}\label{the-shifting-spatial-distribution-of-cycling-and-associated-impacts}

The spatial distribution of cycling potential differs markedly between
scenarios, as illustrated in Figure 7 for the city of Leeds, West
Yorkshire. The top 50 OD pairs in Leeds under Government Target are
strongly influenced by the current distribution of cycling trips,
concentrated in the North of the city (see Figure 3 for comparison with
the baseline). Under the Go Dutch scenario, by contrast, the pattern of
cycling shifts substantially to the South. The cycling patterns under
the Go Dutch scenario are more representative of short-distance trips
across the city overall. In both cases the desire lines are focussed on
Leeds city centre: the region has a mono-centric regional economy,
making commute trips beyond around 5 km from the centre much less likely
to be made by cycling.

\begin{figure}
\includegraphics[width=1\linewidth]{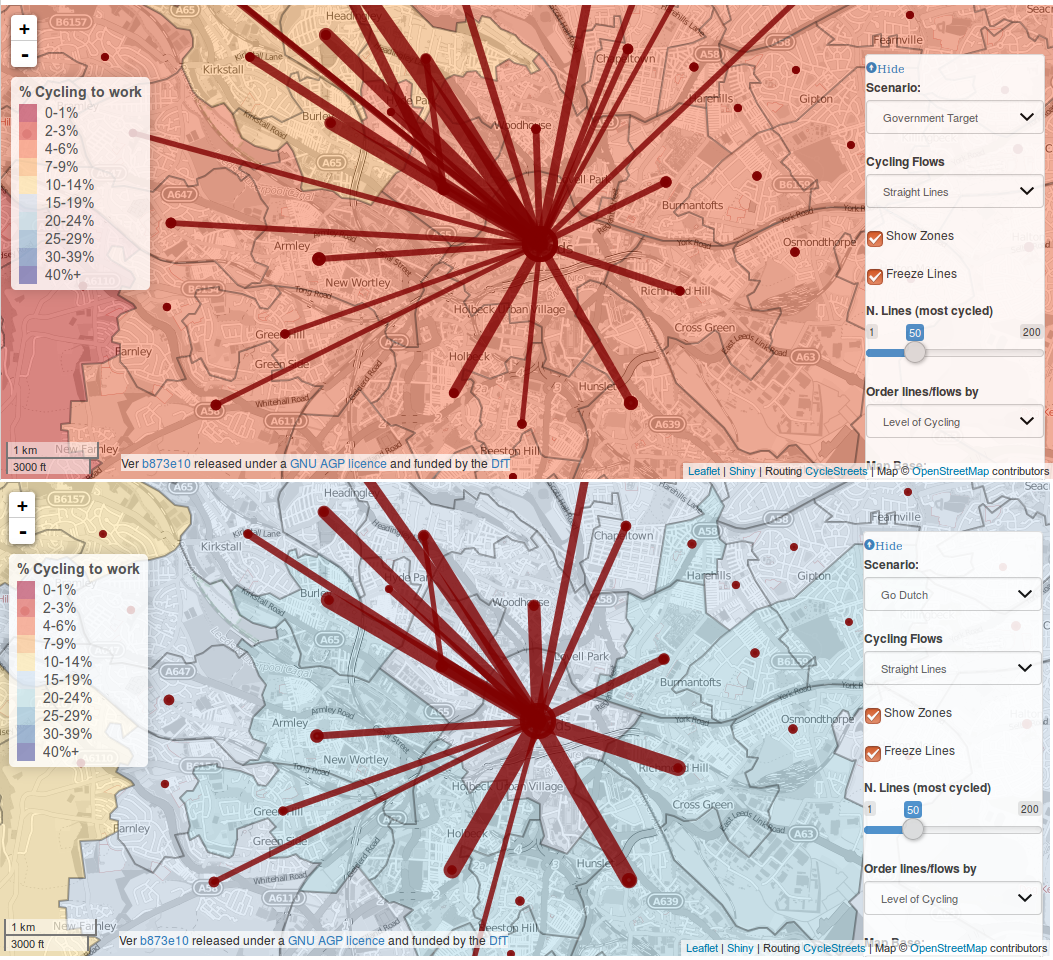} \caption{Model output illustrating the top 50 most cycled OD pairs in Leeds under the Government Target and Go Dutch scenarios.}\label{fig:unnamed-chunk-20}
\end{figure}

\begin{verbatim}
## [1] FALSE
\end{verbatim}

The same scenario is illustrated in Figure 8 with the Route Network
layer. This shows how the number of commuter cyclist using different
road segments could be expected to change. The number using York Road,
highlighted in Figure 8, for example more than triples (from 88 to 318)
under Government Target and increases more than 10 fold under Go Dutch
(from 88 to 1426). This contrasts with Otley Road (highlighted in Figure
5), which `only' triples under the Go Dutch scenario. These outputs
suggest that the geographical distribution of cycling may shift if the
proportion of trips cycled increases in the city. The results also
suggest that cycle paths built to help achieve ambitious targets, as
represented by the Go Dutch scenario, should be of sufficient width to
accommodate the estimated flows.

\begin{figure}
\includegraphics[width=1\linewidth]{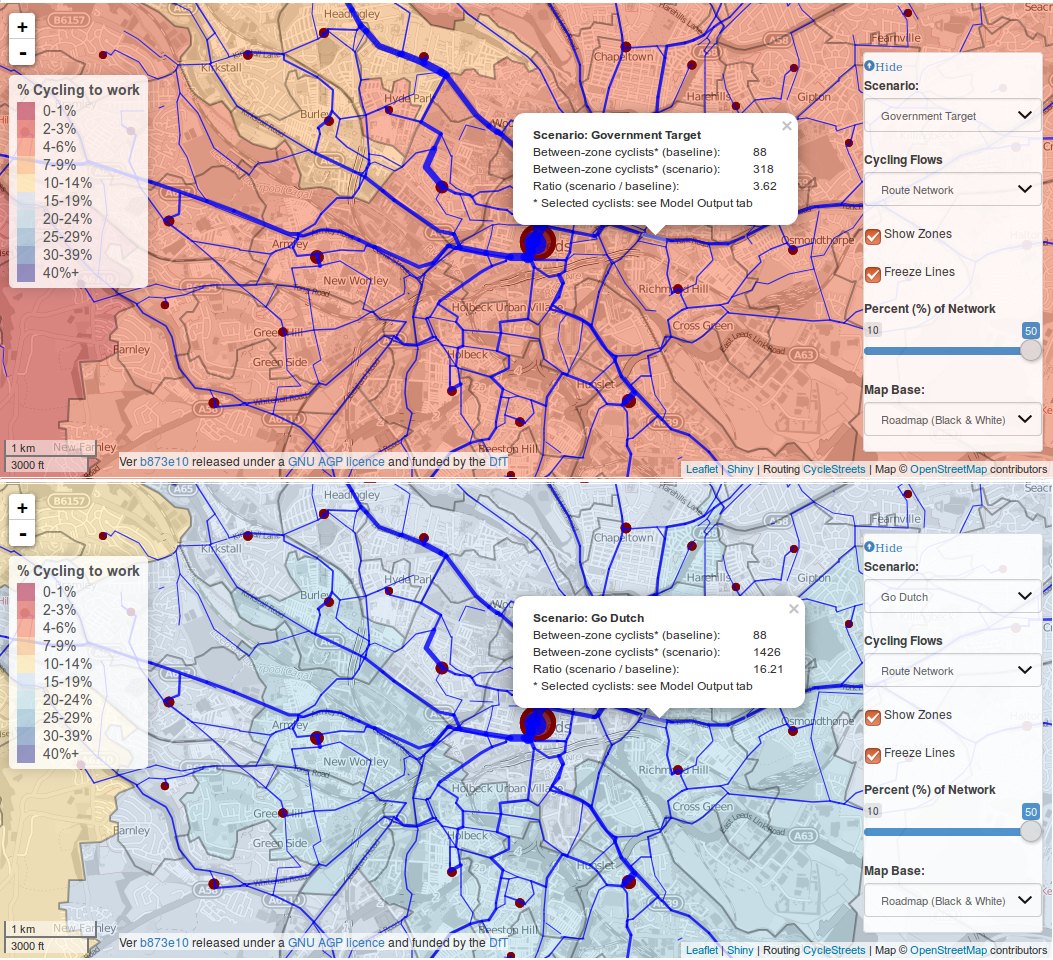} \caption{The Route Network layer illustrating the shifting spatial distribution of cycling flows in Leeds under Government Target (top) and Go Dutch (bottom) scenarios.}\label{fig:unnamed-chunk-21}
\end{figure}

\begin{verbatim}
## [1] FALSE
\end{verbatim}

Another output that can be highly policy relevant is the difference
between `fast' and `quiet' routes. Figure 9 illustrates this by showing
the Fast \& Quiet Routes layer in Manchester with the highest cycling
potential under the Government Target scenario. The `quiet' route is
longer: 2.8 km (as shown by clicking on the line) compared with the more
direct fast route which is 2.3 km. However, the `fast' route may not
currently be attractive for cycling as it passes along a busy dual
carriage way. The Euclidean distance associated with this OD pair is 1.6
km (this can be seen by clicking on a line illustrated from the
`Straight Lines' layer in the PCT's interface), resulting in `circuity'
(see Iacono et al., 2008), values of 1.6 and 1.4 for `quiet' and `fast'
routes respectively.

Dutch guidance suggests that circuity values ``for cycle provision
should be 1.2'' (CROW, 2007). Evidence indicates that women and older
people have a greater preference for off-road and shorter routes
(Garrard et al., 2008; Woodcock et al., 2016). This suggests the `fast
route' option, if built to a high standard, may be favourable from an
equity perspective in this context.

\begin{figure}
\includegraphics[width=1\linewidth]{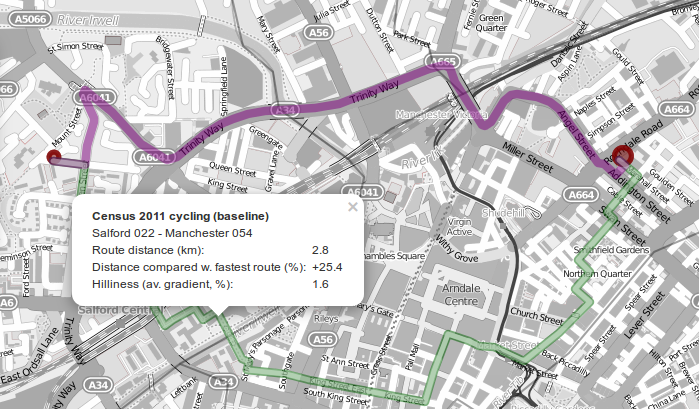} \caption{Close-up of a 'fast' and 'quiet' route in the PCT under the Government Target scenario in Manchester. This provides an indication of the local 'quietness diversion factor'.}\label{fig:unnamed-chunk-22}
\end{figure}

\begin{verbatim}
## [1] FALSE
\end{verbatim}

Three basemap options are worth highlighting in addition to the grey
default basemap. These were selected to provide insight into how the
geographical distribution of latent demand for cycling relates to
current cycle infrastructure and socio-demographics: `OpenCycleMap'
indicates where cycle provision is (and is not) currently; `Index of
Deprivation' illustrates the spatial distribution of social
inequalities; and the `Satellite' basemap can help identify
opportunities for re-allocating space away from roads and other land
uses for cycle and walking paths by providing visual information on road
widths and land uses along desire lines. The `Satellite' basemap option
is illustrated in Figure 10, which shows a section of Trinity Way (as it
crosses the River Irwell). This shows there are 4 lanes of traffic, a
central paved area and wide pavements on both sides of the road,
suggesting there may be space for a cycle path, especially if road space
were re-allocated away from motorised traffic.

\begin{figure}
\includegraphics[width=1\linewidth]{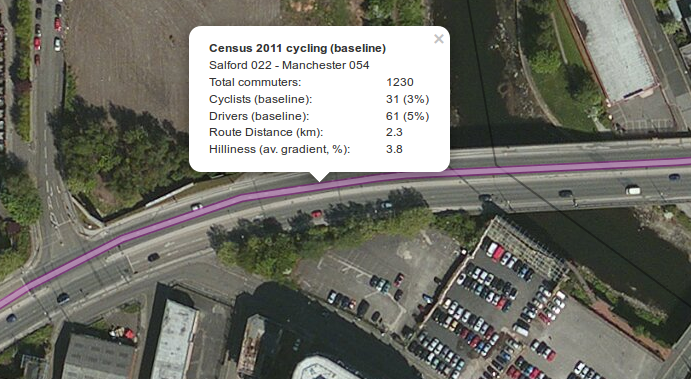} \caption{Zoomed-in section of Trinity way (see Figure 9) using the Satellite basemap to show road width and the number of lanes allocated to motorised traffic.}\label{fig:unnamed-chunk-23}
\end{figure}

\begin{verbatim}
## [1] FALSE
\end{verbatim}

Another feature of the user interface worth highlighting is addition of
a dropdown menu to enable the `top n' routes to be selected not only
based on the level of cycling, but also based on estimated health and
carbon impacts, under each scenario. The reason for this addition was
the finding that health benefits do not always rise in simply in
proportion to the number of people cycling: longer trips lead to a
greater health benefit than short ones do, and this is now represented
by shifting distribution of lines when the ``HEAT Value'' option is
selected from the ``Order lines/flows by'' dropdown menu (this menu only
appears when lines are shown \emph{and} a scenario other than the Census
2011 is selected).

This feature is illustrated in Figure 5, economic value of health
benefits reported for the 4.1 km route in Figure 5 is estimated to be
£70785. When health benefits are the main criteria for policy
evaluation, OD pairs with low current rates of walking would be favoured
for intervention. When emissions are the main criteria, OD pairs with a
high baseline level of car use are also favoured. The exploration of
these considerations is facilitated in the PCT by allowing users to
select the top routes ranked by health and carbon benefits.

\section{Discussion}\label{discussion}

We have outlined a method for modelling and visualising the spatial
distribution of cycling flows, currently and under various scenarios of
`cycling futures'. Inspired by previous approaches to estimating cycling
potential (Larsen et al., 2013; Zhang et al., 2014) and by online,
interactive planning support systems (PSS) (Pettit et al., 2013), the
PCT tackles the issue of how to generate an evidence base to decide
where new cycle paths and other localised pro-cycling interventions
should be prioritised. By showing potential health-related benefits the
tool provides various metrics for transport planners, going beyond the
number of additional trips. Illustrative uses of the PCT demonstrated
the potential utility of the tool, for example by showing settings in
which the spatial distribution of cycling demand is likely to shift as
cycling grows.

In addition to creating an evidence base for planning specific routes
and area-based interventions, the long-term Go Dutch and Ebikes
scenarios could be used for envisioning different transport futures
(Hickman et al., 2011). The PCT could also: help translate national
targets into local aspirations (as illustrated by the Government Target
scenario); inform local targets (e.g.~by indicating what the potential
in one region is relative to neighbouring regions); support business
cases (by showing that there is high cycling potential along proposed
routes); and help plan for cycle capacity along the route network via
the network analysis layer. Ongoing case study work with stakeholders
will be needed to establish and develop these uses. Future developments
will be facilitated by the open source code-base underlying the PCT (see
\href{https://github.com/npct}{github.com/npct}) (Lima et al., 2014).

As with any modelling tool, the approach presented in this paper has
limitations: the reliance on Census origin-destination (OD) data from
2011 means that the results are limited to commuting and may not
encapsulate recent shifts in travel behaviour, and the user interface is
constrained to a few, discrete, scenarios. These limitations suggest
directions for future work, most notably the use of new sources of OD
data.

There is often a tension between transparency and complexity in the
design of tools for transport planning. Excessive complexity can result
in tools that are `black boxes' (Saujot et al., 2016). In a context of
limited time, expertise, and resources, Saujot et al. caution against
investing in ever more complex models. Instead, they suggest models
should be more user focussed. The PCT's open source, freely available
nature will, we believe, facilitate the future development of the PCT
organically to meet the needs of its various users. To encourage others
to use the outputs of the PCT project, we make the data underlying the
online map available to download (via the Zones and Lines tabs in the
tool's online interface). We envision stakeholders in local government
modifying scenarios for their own purposes, and that academics in
relevant fields may add new features and develop new use cases of the
PCT. Such enhancements could include:

\begin{itemize}
\item
  Additional scenarios to illustrate a wider range of `cycling futures',
  including medium-term and local targets.
\item
  Use of individual level data to estimate cycling potential and
  impacts. The use of synthetic `spatial microdata', for example, could
  enable specification of scenarios and analysis of outcomes by a much
  wider range of predictors (Lovelace et al., 2014).
\end{itemize}

\begin{itemize}
\item
  Additional purposes of trips in the model. An `education layer' would
  enable prioritisation of `safe routes to school', building on methods
  analysing `school commute' data (Singleton, 2014). Other data sources
  to include more trip types include mobile telephone providers
  (Alexander et al., 2015) and outputs from transport models.
\item
  Deployment of the PCT for new cities, regions or countries. This will
  depend on the availability of appropriate OD data, perhaps from
  sources mentioned in the previous point, and routing services that can
  estimate cyclable routes based on globally available data, e.g.~using
  the function \texttt{route\_graphhopper} from \textbf{stplanr}
  package. Such work could also facilitate international comparisons of
  cycling potential.
\end{itemize}

Transport planning is a complex and contested field (Banister, 2008).
When it comes to sustainable mobility, policy, politics, leadership and
vision are key ingredients that computer models alone cannot supply
(Melia, 2015). The approach described here can, however, assist in this
wider context by providing new tools for exploring the evidence at high
geographical resolution and envisioning transformational change in
travel behaviours.

By providing transport authorities, campaign groups and the public with
access to the same evidence base, we hypothesise that tools such as the
PCT can encourage informed and rigorous debate, as advocated by Golub et
al. (2013). In conclusion, the PCT provides an accessible evidence base
to inform the question of where to prioritise interventions for active
travel and raises more fundamental questions about how models should be
used in transport planning.

\section*{Author contribution
statement}\label{author-contribution-statement}
\addcontentsline{toc}{section}{Author contribution statement}

The PCT was built as a collaborative effort. The Principal Investigator
of the project was JW, and the initial concept for the project came from
JW and RL in response to a call from the Department of Transport. AG led
the creation of the model underlying the PCT, and the generation of
estimates of cycling levels, health gains and carbon impacts in each
scenario. JW and AG led the development of the methods for calculating
health impacts. JW, RL, AG, and RA contributed to the development of the
cycling uptake rules. RL led the processing of these modelled estimates
for spatial visualisation in the online tool and coordinated the
development of the online tool. RA led the coordination of policy
implementation and collection of practitioner feedback. AA and NB led on
the user interface, with contributions from all authors. NB led the
deployment of the PCT on a public facing server. AG led the writing of
the Appendix. RL led the writing of this manuscript, with input from all
authors.

\section*{Acknowledgements}\label{acknowledgements}
\addcontentsline{toc}{section}{Acknowledgements}

We would like to thank the following people for comments on earlier
versions of the manuscript and the development of the PCT: Roger Geffen
(Cycling UK), Tom Gutowski (Sustrans), Helen Bowkett (Welsh Government),
John Parkin (University of the West of England), Nikée Groot and Phil
Tate. Thanks to Simon Nutall and Martin Lucas-Smith for access to and
instructions on the use of the CycleStreets.net API. Thanks to Barry
Rowlingson from the University of Lancaster for developing functionality
in the \textbf{stplanr} package that enabled the creation of the Route
Network layer. We are grateful to Matthew Tranter at DfT for merging
additional geographic data into the UK National Travel Survey, and to
Eva Heinen, Rick Prins and Thomas Götschi for their help in analysing
Dutch and Swiss Travel Survey data. Thanks to Alvaro Ullrich for
contributions to the code. Thanks to developers of open source software
we have been able to make the PCT free and open to the world. We would
also like to thank Brook Lyndhurst for assistance with the user testing,
and all participants in the user testing sessions. We would also like to
thank Shane Snow and other staff at the DfT for specifying the project's
aims and providing feedback on early versions of the tool.

\section*{Funding}\label{funding}
\addcontentsline{toc}{section}{Funding}

The work presented was funded by the Department for Transport (contract
no. RM5019SO7766: ``Provision of Research Programme into Cycling:
Propensity to Cycle''), with contract facilitation and project
management by Brook Lyndhurst in Phase 1, and by Atkins in Phase 2. RL's
contribution was supported by the Consumer Data Research Centre (ESRC
grant number ES/L011891/1). JW's contribution was supported by an MRC
Population Health Scientist Fellowship. JW's and AA's contributions were
supported by the Centre for Diet and Activity Research (CEDAR), a UKCRC
Public Health Research Centre of Excellence funded by the British Heart
Foundation, Cancer Research UK, Economic and Social Research Council,
Medical Research Council, the National Institute for Health Research
(NIHR), and the Wellcome Trust. AG's contribution was supported by an
NIHR post-doctoral fellowship. The views reported in this paper are
those of the authors and do not necessarily represent those of the DfT,
Brook Lyndhurst, Atkins the NIHR, the NHS or the Department for Health.

\section*{References}\label{references}
\addcontentsline{toc}{section}{References}

\hypertarget{refs}{}
\hypertarget{ref-aldred_does_2016}{}
Aldred, R., Woodcock, J., Goodman, A., 2016. Does More Cycling Mean More
Diversity in Cycling? Transport Reviews 36, 28--44.
doi:\href{https://doi.org/10.1080/01441647.2015.1014451}{10.1080/01441647.2015.1014451}

\hypertarget{ref-alexander_validation_2015}{}
Alexander, L., Jiang, S., Murga, M., Gonz, M.C., 2015. Validation of
origin-destination trips by purpose and time of day inferred from mobile
phone data. Transportation Research Part B: Methodological 1--20.
doi:\href{https://doi.org/10.1016/j.trc.2015.02.018}{10.1016/j.trc.2015.02.018}

\hypertarget{ref-aultman-hall_analysis_1997}{}
Aultman-Hall, L., Hall, F., Baetz, B., 1997. Analysis of bicycle
commuter routes using geographic information systems: Implications for
bicycle planning. Transportation Research Record: Journal of the
Transportation Research Board 1578, 102--110.
doi:\href{https://doi.org/10.3141/1578-13}{10.3141/1578-13}

\hypertarget{ref-banister_sustainable_2008}{}
Banister, D., 2008. The sustainable mobility paradigm. Transport Policy
15, 73--80.
doi:\href{https://doi.org/DOI:\%2010.1016/j.tranpol.2007.10.005}{DOI: 10.1016/j.tranpol.2007.10.005}

\hypertarget{ref-borning_urbansim:_2008}{}
Borning, A., Waddell, P., Förster, R., 2008. UrbanSim: Using simulation
to inform public deliberation and decision-making, in: Digital
Government. Springer, pp. 439--464.

\hypertarget{ref-broach_where_2012}{}
Broach, J., Dill, J., Gliebe, J., 2012. Where do cyclists ride? A route
choice model developed with revealed preference GPS data. Transportation
Research Part A: Policy and Practice 46, 1730--1740.
doi:\href{https://doi.org/10.1016/j.tra.2012.07.005}{10.1016/j.tra.2012.07.005}

\hypertarget{ref-te_brommelstroet_developing_2008}{}
Brömmelstroet, M. te, Bertolini, L., 2008. Developing land use and
transport PSS: Meaningful information through a dialogue between
modelers and planners. Transport Policy 15, 251--259.
doi:\href{https://doi.org/10.1016/j.tranpol.2008.06.001}{10.1016/j.tranpol.2008.06.001}

\hypertarget{ref-buehler_bikeway_2016}{}
Buehler, R., Dill, J., 2016. Bikeway Networks: A Review of Effects on
Cycling. Transport Reviews 36, 9--27.
doi:\href{https://doi.org/10.1080/01441647.2015.1069908}{10.1080/01441647.2015.1069908}

\hypertarget{ref-chan_multi-criteria_2003}{}
Chan, W.-T., Suja, T., 2003. A Multi-Criteria Approach in Designing
Bicycle Tracks, in: Map Asia Conference.

\hypertarget{ref-chatterjee_triggers_2013}{}
Chatterjee, K., Sherwin, H., Jain, J., 2013. Triggers for changes in
cycling: The role of life events and modifications to the external
environment. Journal of Transport Geography 30, 183--193.
doi:\href{https://doi.org/10.1016/j.jtrangeo.2013.02.007}{10.1016/j.jtrangeo.2013.02.007}

\hypertarget{ref-crow_design_2007}{}
CROW, 2007. Design manual for bicycle traffic. Kennisplatform,
Amsterdam.

\hypertarget{ref-de_nazelle_improving_2011}{}
De Nazelle, A., Nieuwenhuijsen, M.J., Antó, J.M., Brauer, M., Briggs,
D., Braun-Fahrlander, C., Cavill, N., Cooper, A.R., Desqueyroux, H.,
Fruin, S., 2011. Improving health through policies that promote active
travel: A review of evidence to support integrated health impact
assessment. Environment international 37, 766--777.

\hypertarget{ref-defra_defra_2015}{}
DEFRA, 2015. DEFRA Carbon Factors: UK Government conversion factors for
Company Reporting, 2015, V2.0.

\hypertarget{ref-department_for_transport_response_2015}{}
Department for Transport, 2015. Response to the consultation on the
draft Cycling Delivery Plan. Department for Transport, London.

\hypertarget{ref-department_for_transport_cycling_2014}{}
Department for Transport, 2014. Cycling Delivery Plan: Draft. Department
for Transport.

\hypertarget{ref-ehrgott_bi-objective_2012}{}
Ehrgott, M., Wang, J.Y.T., Raith, A., Houtte, C. van, 2012. A
bi-objective cyclist route choice model. Transportation Research Part A:
Policy and Practice 46, 652--663.
doi:\href{https://doi.org/10.1016/j.tra.2011.11.015}{10.1016/j.tra.2011.11.015}

\hypertarget{ref-fishman_cycling_2016}{}
Fishman, E., 2016. Cycling as transport. Transport Reviews 36, 1--8.
doi:\href{https://doi.org/10.1080/01441647.2015.1114271}{10.1080/01441647.2015.1114271}

\hypertarget{ref-gaffron_implementation_2003}{}
Gaffron, P., 2003. The implementation of walking and cycling policies in
British local authorities. Transport Policy 10, 235--244.
doi:\href{https://doi.org/10.1016/S0967-070X(03)00024-6}{10.1016/S0967-070X(03)00024-6}

\hypertarget{ref-garrard_promoting_2008}{}
Garrard, J., Rose, G., Lo, S.K., 2008. Promoting transportation cycling
for women: The role of bicycle infrastructure. Preventive Medicine 46,
55--59.
doi:\href{https://doi.org/10.1016/j.ypmed.2007.07.010}{10.1016/j.ypmed.2007.07.010}

\hypertarget{ref-golub_making_2013}{}
Golub, A., Robinson, G., Nee, B., 2013. Making accessibility analyses
accessible: A tool to facilitate the public review of the effects of
regional transportation plans on accessibility. Journal of Transport and
Land Use 6, 17--28.
doi:\href{https://doi.org/10.5198/jtlu.v6i3.352}{10.5198/jtlu.v6i3.352}

\hypertarget{ref-haklay_web_2008}{}
Haklay, M., Singleton, A., Parker, C., 2008. Web Mapping 2.0: The
Neogeography of the GeoWeb. Geography Compass 2, 2011--2039.
doi:\href{https://doi.org/10.1111/j.1749-8198.2008.00167.x}{10.1111/j.1749-8198.2008.00167.x}

\hypertarget{ref-han_assessment_2008}{}
Han, J., Hayashi, Y., 2008. Assessment of private car stock and its
environmental impacts in China from 2000 to 2020. Transportation
Research Part D: Transport and Environment 13, 471--478.

\hypertarget{ref-heath_effectiveness_2006}{}
Heath, G.W., Brownson, R.C., Kruger, J., Miles, R., Powell, K.E.,
Ramsey, L.T., Services, T.F. on C.P., 2006. The effectiveness of urban
design and land use and transport policies and practices to increase
physical activity: A systematic review. Journal of Physical Activity \&
Health 3, S55.

\hypertarget{ref-heinen_changes_2015}{}
Heinen, E., Panter, J., Mackett, R., Ogilvie, D., 2015. Changes in mode
of travel to work: A natural experimental study of new transport
infrastructure. International Journal of Behavioral Nutrition and
Physical Activity 12, 81.
doi:\href{https://doi.org/10.1186/s12966-015-0239-8}{10.1186/s12966-015-0239-8}

\hypertarget{ref-hickman_transitions_2011}{}
Hickman, R., Ashiru, O., Banister, D., 2011. Transitions to low carbon
transport futures: Strategic conversations from London and Delhi.
Journal of Transport Geography, Special section on Alternative Travel
futures 19, 1553--1562.
doi:\href{https://doi.org/10.1016/j.jtrangeo.2011.03.013}{10.1016/j.jtrangeo.2011.03.013}

\hypertarget{ref-iacono_access_2008}{}
Iacono, M., Krizek, K., El-Geneidy, A., 2008. Access to Destinations:
How Close is Close Enough? Estimating Accurate Distance Decay Functions
for Multiple Modes and Different Purposes 76.

\hypertarget{ref-kahila-tani_let_2016}{}
Kahila-Tani, M., Broberg, A., Kyttä, M., Tyger, T., 2016. Let the
Citizens Map---Public Participation GIS as a Planning Support System in
the Helsinki Master Plan Process. Planning Practice \& Research 31,
195--214.
doi:\href{https://doi.org/10.1080/02697459.2015.1104203}{10.1080/02697459.2015.1104203}

\hypertarget{ref-kahlmeier_health_2014}{}
Kahlmeier, S., Kelly, P., Foster, C., Götschi, T., Cavill, N., Dinsdale,
H., Woodcock, J., Schweizer, C., Rutter, H., Lieb, C., 2014. Health
economic assessment tools (HEAT) for walking and for cycling, Methods
and User Guide. World Health Organization Regional Office for Europe,
Copenhagen, Denmark 2014.

\hypertarget{ref-klosterman_what_1999}{}
Klosterman, R.E., 1999. The What If? Collaborative Planning Support
System. Environment and Planning B: Planning and Design 26, 393--408.
doi:\href{https://doi.org/10.1068/b260393}{10.1068/b260393}

\hypertarget{ref-komanoff_bicycling_2004}{}
Komanoff, C., 2004. Bicycling, in: Cleveland, C.J. (Ed.), Encyclopedia
of Energy. Elsevier, New York, pp. 141--150.

\hypertarget{ref-kuzmyak_estimating_2014}{}
Kuzmyak, J.R., Walters, J., Bradley, M., Kockelman, K., 2014. Estimating
bicycling and walking for planning and project development, Nchrp
national cooperative highway research program report. Transportation
Research Board of the National Academies, Washington, DC.

\hypertarget{ref-larsen_build_2013}{}
Larsen, J., Patterson, Z., El-Geneidy, A., 2013. Build it. But where?
The use of geographic information systems in identifying locations for
new cycling infrastructure. International Journal of Sustainable
Transportation 7, 299--317.

\hypertarget{ref-lima_coding_2014}{}
Lima, A., Rossi, L., Musolesi, M., 2014. Coding Together at Scale:
GitHub as a Collaborative Social Network. arXiv preprint
arXiv:1407.2535.

\hypertarget{ref-lovelace_spatial_2014}{}
Lovelace, R., Ballas, D., Watson, M., 2014. A spatial microsimulation
approach for the analysis of commuter patterns: From individual to
regional levels. Journal of Transport Geography 34, 282--296.
doi:\href{https://doi.org/http://dx.doi.org/10.1016/j.jtrangeo.2013.07.008}{http://dx.doi.org/10.1016/j.jtrangeo.2013.07.008}

\hypertarget{ref-lovelace_assessing_2011}{}
Lovelace, R., Beck, S., Watson, M., Wild, A., 2011. Assessing the energy
implications of replacing car trips with bicycle trips in Sheffield, UK.
Energy Policy 39, 2075--2087.
doi:\href{https://doi.org/10.1016/j.enpol.2011.01.051}{10.1016/j.enpol.2011.01.051}

\hypertarget{ref-mccollum_achieving_2009}{}
McCollum, D., Yang, C., 2009. Achieving deep reductions in US transport
greenhouse gas emissions: Scenario analysis and policy implications.
Energy Policy 37, 5580--5596.
doi:\href{https://doi.org/10.1016/j.enpol.2009.08.038}{10.1016/j.enpol.2009.08.038}

\hypertarget{ref-melia_urban_2015}{}
Melia, S., 2015. Urban Transport Without the Hot Air, Volume 1:
Sustainable Solutions for UK cities.

\hypertarget{ref-minikel_cyclist_2012}{}
Minikel, E., 2012. Cyclist safety on bicycle boulevards and parallel
arterial routes in Berkeley, California. Accident Analysis \& Prevention
45, 241--247.

\hypertarget{ref-novosel_agent_2015}{}
Novosel, T., Perković, L., Ban, M., Keko, H., Pukšec, T., Krajačić, G.,
Duić, N., 2015. Agent based modelling and energy planning--Utilization
of MATSim for transport energy demand modelling. Energy.

\hypertarget{ref-oja_health_2011}{}
Oja, P., Titze, S., Bauman, A., DeGeus, B., Krenn, P., Reger-Nash, B.,
Kohlberger, T., 2011. Health benefits of cycling: A systematic review.
Scandinavian journal of medicine \& science in sports 21, 496--509.
doi:\href{https://doi.org/10.1111/j.1600-0838.2011.01299.x}{10.1111/j.1600-0838.2011.01299.x}

\hypertarget{ref-parkin_planning_2015}{}
Parkin, J., 2015. Planning and design approaches for cycling
infrastructure.

\hypertarget{ref-parkin_cycling_2012}{}
Parkin, J. (Ed.), 2012. Cycling and sustainability, 1. ed. ed, Transport
and sustainability. Emerald, Bingley.

\hypertarget{ref-parkin_estimation_2008}{}
Parkin, J., Wardman, M., Page, M., 2008. Estimation of the determinants
of bicycle mode share for the journey to work using census data.
Transportation 35, 93--109.
doi:\href{https://doi.org/10.1007/s11116-007-9137-5}{10.1007/s11116-007-9137-5}

\hypertarget{ref-payne_removing_2014}{}
Payne, S., 2014. Removing barriers to direct access. Get Britain Cycling
4, 6--8.

\hypertarget{ref-pettit_online_2013}{}
Pettit, C.J., Klosterman, R.E., Nino-Ruiz, M., Widjaja, I., Russo, P.,
Tomko, M., Sinnott, R., Stimson, R., 2013. The online what if? Planning
support system, in: Planning Support Systems for Sustainable Urban
Development. Springer, pp. 349--362.

\hypertarget{ref-pikora_developing_2002}{}
Pikora, T.J., Bull, F.C.L., Jamrozik, K., Knuiman, M., Giles-Corti, B.,
Donovan, R.J., 2002. Developing a reliable audit instrument to measure
the physical environment for physical activity. American Journal of
Preventive Medicine 23, 187--194.
doi:\href{https://doi.org/10.1016/S0749-3797(02)00498-1}{10.1016/S0749-3797(02)00498-1}

\hypertarget{ref-pucher_infrastructure_2010}{}
Pucher, J., Dill, J., Handy, S., 2010. Infrastructure, programs, and
policies to increase bicycling: An international review. Preventive
Medicine 50, S106--S125.
doi:\href{https://doi.org/DOI:\%2010.1016/j.ypmed.2009.07.028}{DOI: 10.1016/j.ypmed.2009.07.028}

\hypertarget{ref-rae_spatial_2009}{}
Rae, A., 2009. From spatial interaction data to spatial interaction
information? Geovisualisation and spatial structures of migration from
the 2001 UK census. Computers, Environment and Urban Systems 33,
161--178.
doi:\href{https://doi.org/10.1016/j.compenvurbsys.2009.01.007}{10.1016/j.compenvurbsys.2009.01.007}

\hypertarget{ref-saujot_making_2016}{}
Saujot, M., Lapparent, M. de, Arnaud, E., Prados, E., 2016. Making land
use -- Transport models operational tools for planning: From a top-down
to an end-user approach. Transport Policy 49, 20--29.
doi:\href{https://doi.org/10.1016/j.tranpol.2016.03.005}{10.1016/j.tranpol.2016.03.005}

\hypertarget{ref-shergold_rural_2012}{}
Shergold, I., Parkhurst, G., Musselwhite, C., 2012. Rural car
dependence: An emerging barrier to community activity for older people.
Transportation Planning and Technology 35, 69--85.

\hypertarget{ref-singleton_gis_2014}{}
Singleton, A., 2014. A GIS approach to modelling CO2 emissions
associated with the pupil-school commute. International Journal of
Geographical Information Science 28, 256--273.
doi:\href{https://doi.org/10.1080/13658816.2013.832765}{10.1080/13658816.2013.832765}

\hypertarget{ref-tainio_can_2016}{}
Tainio, M., Nazelle, A.J. de, Götschi, T., Kahlmeier, S., Rojas-Rueda,
D., Nieuwenhuijsen, M.J., Sá, T.H. de, Kelly, P., Woodcock, J., 2016.
Can air pollution negate the health benefits of cycling and walking?
Preventive Medicine.
doi:\href{https://doi.org/10.1016/j.ypmed.2016.02.002}{10.1016/j.ypmed.2016.02.002}

\hypertarget{ref-tamminga_design_2012}{}
Tamminga, G., Miska, M., Santos, E., Lint, H. van, Nakasone, A.,
Prendinger, H., Hoogendoorn, S., 2012. Design of Open Source Framework
for Traffic and Travel Simulation. Transportation Research Record:
Journal of the Transportation Research Board 2291, 44--52.
doi:\href{https://doi.org/10.3141/2291-06}{10.3141/2291-06}

\hypertarget{ref-tobler_experiments_1987}{}
Tobler, W.R., 1987. Experiments in migration mapping by computer. The
American Cartographer 14, 155--163.

\hypertarget{ref-transport_for_london_london_2015}{}
Transport for London, 2015. London Cycling Design Standards (LCDS).

\hypertarget{ref-transport_for_london_analysis_2010}{}
Transport for London, 2010. Analysis of Cycling Potential. Transport for
London.

\hypertarget{ref-wegman_urban_1979}{}
Wegman, F.C.M., 1979. Urban planning, traffic planning and traffic
safety of pedestrians and cyclists, in: Contribution to the
OECD-Symposium in Safety of Pedestrians and Cyclists, Paris.

\hypertarget{ref-welsh_government_guidance_2014}{}
Welsh Government, 2014. Guidance Active Travel (Wales) Act. Welsh
Government, Cardiff.

\hypertarget{ref-wood_visualisation_2010}{}
Wood, J., Dykes, J., Slingsby, A., 2010. Visualisation of origins,
destinations and flows with OD maps. The Cartographic Journal 47,
117--129.

\hypertarget{ref-woodcock_national_2016}{}
Woodcock, J., Aldred, R., Goodman, A., Lovelace, R., Ullrich, A., Abbas,
A., Berkoff, N., 2016. National Propensity to Cycle Tool Project:
Summary Report. Department for Transport.

\hypertarget{ref-woodcock_public_2009}{}
Woodcock, J., Edwards, P., Tonne, C., Armstrong, B.G., Ashiru, O.,
Banister, D., Beevers, S., Chalabi, Z., Chowdhury, Z., Cohen, A.,
Franco, O.H., Haines, A., Hickman, R., Lindsay, G., Mittal, I., Mohan,
D., Tiwari, G., Woodward, A., Roberts, I., 2009. Public health benefits
of strategies to reduce greenhouse-gas emissions: Urban land transport.
The Lancet 374, 1930--1943.
doi:\href{https://doi.org/10.1016/S0140-6736(09)61714-1}{10.1016/S0140-6736(09)61714-1}

\hypertarget{ref-zhang_prioritizing_2014}{}
Zhang, D., Magalhaes, D.J.A.V., Wang, X.(., 2014. Prioritizing bicycle
paths in Belo Horizonte City, Brazil: Analysis based on user preferences
and willingness considering individual heterogeneity. Transportation
Research Part A: Policy and Practice 67, 268--278.
doi:\href{https://doi.org/10.1016/j.tra.2014.07.010}{10.1016/j.tra.2014.07.010}

\end{document}